\documentstyle[12pt,epsfig]{article}
\newcommand{\ba}{\begin{array}}
\newcommand{\ea}{\end{array}}
\newcommand{\bd}{\begin{displaymath}}
\newcommand{\ed}{\end{displaymath}}
\newcommand{\be}{\begin{equation}}
\newcommand{\ee}{\end{equation}}
\newcommand{\bea}{\begin{eqnarray}}
\newcommand{\eea}{\end{eqnarray}}

\def\barr{\begin{array}}
\def\earr{\end{array}}

\newcommand{\beq}{\begin{eqnarray}}
\newcommand{\eqn}{\end{eqnarray}}

\def\etal{ {\em et al.}}

\def\q2 {q^2}

\def\N10{\widetilde \chi_1^0}
\def\Cp1{\widetilde \chi_1^+}
\def\Cm1{\widetilde \chi_1^-}
\def\C1pm{\widetilde \chi_1^\pm}

\begin{document}
\begin{titlepage}

\begin{flushleft}
%\today\\
%version-4
\end{flushleft}
\begin{flushright}
{\large OITS-742}
\end{flushright}

\begin{center}
{\Large \bf Flavor and CP violating $Z$ exchange and the rate asymmetry in $B \to \phi K_S $}\\[10mm]
{\large N.G. Deshpande\footnote{desh@OREGON.UOREGON.EDU}
and Dilip Kumar Ghosh\footnote{dghosh@physics.uoregon.edu}} \\[4mm]
Institute of Theoretical Science \\
University of Oregon, Eugene, OR 97403\\[10mm]
\end{center}

\begin{abstract}
\noindent
Recent measurements of time dependent CP asymmetry in $B \to \phi K_S$, if 
confirmed, would indicate a new source of CP violation. We examine flavor 
violating tree-level $Z$ currents in models with extra down-type 
quark singlets that arise naturally in string compactified gauge groups 
like $E_6$. We evaluate the new operators at the scale 
$\mu \approx {\cal O} (m_b) $ in NLO,
and using QCD improved factorization to describe $B \to \phi K_S $, find the 
allowed range of parameters for $\rho$ and $\psi$, the magnitude and phase 
of the flavor violating parameter $U_{bs}$. This allowed range does 
satisfy the constraint from flavor changing process $b \to s \ell^+\ell^-$.
However, further improvement in measurement of these rates could severely 
constrain the model.
\end{abstract}
%PACS number(s): 13.25.Hw, 13.20.He, 12.38.Bx, 12.60.J

\end{titlepage}

\section{Introduction}
The ongoing $B$ physics experiments by BaBar and Belle collaborations~
\cite{babar, belle1} provide a unique opportunity to study the flavor 
structure of the standard model quark sector and also the origin of CP 
violation. In addition to this, any new physics effects in $B$ physics 
can also be tested in these experiments. 
Recent time dependent asymmetries measured in the decay 
$B \to \phi K_S$ both by 
BaBar and Belle collaborations [\ref{babar}-\ref{hamel}]
%\cite{babar, belle1, belle2,hamel} 
show 
significant deviation 
from the standard model and this has generated much theoretical speculation 
regarding physics beyond the standard model \cite{np1}. In the standard
model, the process $B \to \phi K_S$ is purely penguin dominated and the
leading contribution has no weak phase. The coefficient of 
$\sin(\Delta m_B t) $ in the asymmetry therefore should measure 
$\sin 2\beta $, the same quantity that is involved in $B \to \psi K_S$ 
in the standard model. The most recent measured average values of 
asymmetries are ~\cite{hamel,browder}
$ S_{\psi K_S}  =  0.734 \pm 0.055 $ and  
$ S_{\phi K_S}  =  -0.15 \pm 0.33  $.
The value for $ S_{\psi K_S} $ agrees with the standard model expectation.
The deviation in the $\phi K_S$ is intriguing because a penguin process 
being a loop induced process is particularly sensitive to new physics which 
can manifest itself through exchange of heavy particles. 
In this article we will consider an extension of the standard model, with 
extra down type singlet quarks. These extra down type singlet quarks
appear naturally in each 27-plet fermion generation of $E_6$ Grand Unification
Theories (GUTs) [\ref{jlr}-\ref{bbp}].
%\cite{jlr, bdpw,rizzo, bbp}. 
The mixing of these singlet 
quarks with with the three SM down type quarks, provides a framework to
study the deviations from the unitarity constraints of $3\times 3 $ CKM
matrix. This model has been previously studied in connection with $R_b$ and 
F-B asymmetry at the $Z$ pole as it provides a framework for violation of 
the unitarity of the CKM matrix [\ref{bbp}-\ref{aguila}].
%\cite{bbp,b_rev1,aguila}. 
This mixing also induces tree-level flavor changing neutral
currents (FCNC). These tree-level FCNC couplings can have a 
significant effects on different $CP$ conserving as well as CP 
violating B processes [\ref{brev1}, \ref{xyz1} - \ref{xyz2}].
%\cite{b_rev1,nagashima,chang-chang-keung,
%botella1, botella2, hiller,dh_ds,jaas, giri}.

In this article we study the FCNC 
effect arising from the $Z-b-\bar s$ coupling $U_{bs}$ to the  
$B\to \phi K_S$ process. This new FCNC coupling $U_{bs}$ can have a phase,
which can generate the additional source of CP violation in the 
$B \to \phi K_S$ process, and thus affect measured values of $S_{\phi K_S}$
and $C_{\phi K_S}$. 
We parameterize this coupling by $U_{bs} = \rho e^{i\psi} $. 
We then study $B \to \phi K_S$ taking into account the new interactions 
in the QCD improved factorization scheme (BBNS approach ) \cite {beneke}. 
This method incorporates elements of
naive factorization approach (as its leading term ) and perturbative QCD 
corrections (as sub-leading contributions) and 
allows one to compute systematic
radiative corrections to the naive factorization for the hadronic $B$ decays.
Recently, several studies of $B \to PV$, and specifically $B \to \phi K_S $ 
have been performed within the frame work of QCD improved factorization 
scheme~[\ref{cheng-yang}-\ref{jpma}]. 
%\cite{cheng-yang,du1,du2,alekson,jpma} 
In our analysis of $B\to \phi K_S$, we follow \cite{jpma} which is based
on the original paper \cite{beneke}. In our analysis, we only
consider the contribution of the leading twist meson wave functions, and also
neglect the weak annihilation contribution which is expected to be small. 
Inclusion of these would introduce more
model dependence in the calculation through the parameterization of an 
integral, which is otherwise infrared divergent.      

The time dependent CP asymmetry of $B \to \phi K_S$ is described by :
\beq
{\cal A}_{\phi K_S}(t) &=& \frac{\Gamma(\overline{ B^0}(t) \to \phi K_S) 
- \Gamma (B^0(t) \to \phi K_S)}{ \Gamma(\overline{ B^0}(t) \to \phi K_S)+
\Gamma (B^0(t) \to \phi K_S)}\\
&=& -C_{\phi K_S} \cos (\Delta m_B t ) + S_{\phi K_S} \sin(\Delta m_B t )
\eqn
where $S_{\phi K_S}$ and $C_{\phi K_S} $ are given by 

\beq
S_{\phi K_S} = \frac{2 Im~\lambda_{\phi K_S}}{1 + \mid \lambda_{\phi K_S}\mid^2}
,~~~~ 
C_{\phi K_S} = \frac{1- \mid \lambda_{\phi K_S}\mid^2}
{1 + \mid \lambda_{\phi K_S}\mid^2}
\eqn
and $\lambda_{\phi K_S}$ can be expressed in terms of decay amplitudes:
\beq
\lambda_{\phi K_S} = -e^{-2i \beta }\frac{{\overline {\cal M}}(\overline{B^0 }\to \phi K_S)}{{\cal M}(B^0 \to \phi K_S)}
\eqn

The branching ratio and the direct CP asymmetries of both the charged and 
neutral modes of $B \to \phi K_S$ have been measured
\cite{babar, belle1, belle2,hamel,browder,CLEO}
\footnote{Latest results 
were reported at XXXIX Rencontres de Moriond, Electroweak Interactions and
Unified Theories, Italy, March 2004. See talks \cite{ccwang,mverderi,nakamura}.
The Belle result on $S_{\phi K}$ is unchanged \cite{ccwang} while 
BaBar finds $S_{\phi K} = 0.47\pm 0.34^{+0.08}_{-0.06} $ \cite{mverderi} 
which is very close to the result in Eq.(7). Hence, our observations remain
unchanged.}
\beq
{\cal B}(B^0 \to \phi K_S) & =& (8.0\pm 1.3) \times 10^{-6}\\
{\cal B}(B^+ \to \phi K^+) & =& (9.4 \pm 0.9) \times 10^{-6} ,\\
S_{\phi K_S} & = & +0.45\pm 0.43 \pm 0.07~~(\rm {BaBar}); \\ 
             & = & -0.96\pm 0.50^{+0.09}_{-0.11}~~({\rm Belle });\\
             & = & -0.15\pm 0.33~~({\rm World~average });\\
C_{\phi K_S} & = & -0.19 \pm 0.30 \\
{\cal A}_{CP}(B^+ \to \phi K^+) & = & (3.9 \pm 8.8 \pm 1.1)\% 
\eqn
\begin{table}
\begin{center}
\begin{tabular}{|c|c|c|c|c|c|c|}
\hline 
scale & $C_1 $ & $C_2 $ & $C_3 $ & $C_4 $& $C_5 $ & $C_6$ \\
\hline 
$\mu = m_b/2 $ & 1.137 & -0.295 & 0.021 & -0.051 & 0.010 & -0.065 \\
$\mu = m_b $ & 1.081 & -0.190 & 0.014 & -0.036 & 0.009 & -0.042 \\
\hline 
        &  $C_7/\alpha_{em} $ & $C_8/\alpha_{em} $ & $C_9/\alpha_{em} $ 
& $C_{10}/\alpha_{em} $
& $C_{7\gamma} $ & $C_{8g}$ \\
\hline 
$\mu = m_b/2 $ & -0.024 & 0.096 & -1.325 & 0.331 & -0.364  & -0.169 \\
\hline 
$\mu = m_b $ & -0.011 & 0.060 & -1.254 & 0.223 & -0.318 & -0.151 \\
\hline
\hline
\end{tabular}
\caption{Standard model Wilson coefficients in NDR scheme.}
\end{center}
\end{table}

\section{ $B \to \phi K_S $ in the QCDF Approach}
In the standard model, the effective Hamiltonian for charmless 
$B \to \phi K_S $ decay is given by \cite{beneke} 
\beq
{\cal H}_{eff} = -\frac{G_F}{\sqrt{2}} V_{tb}V^{\ast}_{ts}
\left [ C_1(\mu) {\cal O}_1(\mu) + C_2(\mu) {\cal O}_2(\mu) 
+ \sum^{10}_{i= 3} C_i(\mu) {\cal O}_i(\mu)  + C_{7 \gamma} {\cal O}_{7 \gamma}
+ C_{8 g} {\cal O}_{8 g}\right ]
\eqn 
where the Wilson coefficients $C_i(\mu)$ are obtained from the weak scale 
down to scale $\mu$ by running the renormalization group equations. The 
definitions of the operators can be found in Ref.\cite{beneke}.
The Wilson coefficients $C_i$ can be computed using different schemes
\cite{wilson_coeff}. In this paper we will use the NDR scheme. The NLO 
values of $C_i (i =1-10)$ and LO values of 
$C_{7\gamma}, C_{8g}$ respectively at $\mu = m_b/2$ and $ m_b $ used by 
us based on Ref.\cite{beneke} are shown in Table~1. 

In the QCD improved factorization scheme, the $B \to \phi K_S $ decay amplitude 
due to a particular operator can be represented in following form :
\beq
< \phi K \mid {\cal O}\mid B> = < \phi K \mid {\cal O}\mid B>_{fact}
\left [1 + \sum r_n \alpha_{s}^n + O(\Lambda_{QCD}/m_b)\right ] 
\eqn
where $< \phi K \mid {\cal O}\mid B>_{fact}$ denotes the naive factorization 
result.The second and third term in the bracket represent higher order 
$\alpha_s$ and $ \Lambda_{QCD}/m_b $ correction to the hadronic transition
amplitude. Following the scheme and notations presented in 
Ref.\cite{jpma}, we write down the total $B \to \phi K_S $ amplitude, which is
the sum of the standard model as well as $Z$ exchange tree-level contribution
from extra down-type quark singlets (EDQS) model in the heavy quark limit 
\beq
{\cal M }(B^+ \to \phi K^+ ) &=&
\nonumber \\  
{\cal M }(B^0 \to \phi K^0 )
&=&\frac{G_F}{\sqrt{2}}m_B^2 f_{\phi} F_1^{B \to K}(m^2_{\phi})
V_{pb}V^{\ast}_{ps}\left [ a^p_3 + a^p_4+ a^p_5 
\right.
\nonumber \\ 
&&
\left.
-\frac{(a^p_7+a^p_9+a^p_{10})}{2} + a^p_{10a} \right ]
\eqn
where $p$ is summed over $u$ and $c$. The coefficients $a^p_i$ are given by 
\begin{eqnarray}
&&a^u_3 = a^c_3 =  C'_3 + {C'_4 \over N_c}
\left[ 1 + {C_F \alpha_s \over 4 \pi} \left (V_{\phi} + H_{\phi}\right )\right],
\nonumber\\ 
&&a^p_4  = 
C'_4 + {C'_3\over N_c} \left[ 1 + {C_F \alpha_s \over 4 \pi} 
\left (V_{\phi} + H_{\phi}\right ) \right] + 
{C_F \alpha_s \over 4\pi N_c} P^p_{\phi}, 
\nonumber\\ 
&&a^u_5  = a^c_5=  C'_5 + {C'_6 \over N_c}
\left[ 1 + {C_F \alpha_s \over 4 \pi} 
(-12-V_{\phi}) \right],
\nonumber\\ 
&&a^u_7  =  a^c_7 = C'_7 + {C'_8 \over N_c}
\left[ 1 + {C_F \alpha_s \over 4 \pi} 
(-12-V_{\phi} - H_{\phi}) \right],
\nonumber\\ 
&&a^u_9  = a^c_9 =  C'_9 + {C'_{10} \over N_c}
\left[ 1 + {C_F \alpha_s \over 4 \pi} 
\left (V_{\phi} + H_{\phi}\right ) \right],
\nonumber\\ 
&&a^u_{10}  =  a^c_{10} = C'_{10} + {C'_9 \over N_c}
\left[ 1 + {C_F \alpha_s \over 4 \pi} \left (V_{\phi} + H_{\phi}\right )
\right],
\nonumber \\
&& a^p_{10} =  {C_F \alpha_s \over 4 \pi  N_c} Q^p_{\phi}
\nonumber \\
&& a^u_{10a} = a^c_{10a}= {C_F \alpha_s \over 4 \pi  N_c} Q_{\phi}
\end{eqnarray}
with $C_F = (N^2_c-1)/2N_c$ and $N_c = 3$. The effective Wilson coefficients
$C^\prime_i = C_i + \tilde C_i,~~~(i =3-10)$ is the sum of the standard model
and the EDQS model Wilson coefficients.
The quantities $V_{\phi}, H_{\phi}, P^p_{\phi}$ and $Q^p_{\phi}$ are hadronic
parameters that contain all nonperturbative dynamics, are given in 
Ref. \cite{beneke, desh_dkg}.

For the sake of completeness, we give the branching ratio for $B\to \phi K_S$ 
decay channel in the rest frame of the $B$ meson.
\beq
{\cal BR}(B\to \phi K_S) = \frac{\tau_B}{8 \pi}\frac{\mid P_{cm} \mid}{m^2_B}
\mid {\cal M}(B \to \phi K_S) \mid^2  
\eqn
where, $\tau_B$ represents the $B$ meson lifetime and the kinematical factor
$\mid P_{cm}\mid $ is written as 
\beq
\mid P_{cm} \mid = \frac{1}{2 m_B}\sqrt{\left [m^2_B - (m_K+m_\phi)^2\right]
\left [m^2_B -(m_K - m_\phi)^2\right] } 
\eqn

\section{FCNC $Z$ couplings in EDQS model}
Models with extra down-type quarks (EDQS) have a long history. The earliest 
consideration of such models was in the context of the grand unification 
group $E_6$ which arises from string compactification. The quarks and leptons
of each generation belong to the $27$ representation [\ref{jlr}-\ref{bbp}].
%\cite{jlr,bdpw,rizzo,bbp}.
Each generation has one extra quark singlet of the down type, and also one 
extra lepton of the 
electron type. The group also has extra $Z$ bosons, which we will assume to be
too heavy to have any effect on $B \to \phi K_S$ process. The down type mass 
matrix is then a $6\times 6$ by-unitary matrix, and in general when we rotate 
the quarks to their mass basis, off-diagonal couplings arise. 
In EDQS model, the $Z$ mediated FCNC interactions are given by
\beq
{\cal L} = \frac{g}{2 \cos \theta_W}\left [
\bar d_{L \alpha } U_{\alpha \beta} \gamma^\mu d_{L\beta }\right] Z_{\mu} 
\eqn
In general for 
$n$ copies of extra down-type quark singlet model, $U_{\alpha\beta}$
is :
\beq
U_{\alpha \beta } = \sum^{3}_{i = 1} V^\dagger_{\alpha i} V_{i \beta} = 
\delta_{\alpha\beta} - \sum^{N_d}_{i = 4} V^{\dagger}_{\alpha i } V_{i\beta},~~~~~
(\alpha, \beta =d,s,b,B_1,B_2....)
\label{uij}
\eqn
where, $N_d= 3 + n $ represents the number of down type quark states,
and $U$ is the neutral current mixing matrix 
for the down quark sector. 
%For a single generation of extra 
%down-type quark singlet, Eq.~(\ref{uij}) can be simplified to :
%$U_{\alpha\beta} = -V^{*}_{4\alpha} V_{4\beta} \neq 0 
%$ for $\alpha \neq \beta $. For a single generation of 
%extra down type quarks singlet, from Eq.~(\ref{uij}) one can get following 
%relation \cite{botella1} 
%\beq
%U_{bd} U^{*}_{bs} = - U_{sd} \mid V_{4b}\mid^2
%\label{bd_bs_rel}
%\eqn
%However, in our analysis we will consider the most general extra down-type 
%quark singlet model (n EDQS), for reason explained later. 
The non vanishing components
of $U_{\alpha\beta}$ will lead to FCNC process
at tree level, generating new physics contribution to the measured CP 
asymmetries. The new tree level FCNC $Z$ mediated contribution 
to the $ b \to s q \bar q$ process is shown in Figure 1. The new operators
arising from this tree-level FCNC process have been shown to lead to the
following effective Hamiltonian for $b \to s q \bar q$ process in this model
~\cite{atwood_hiller}:
%\beq 
%{\cal O}^{\prime}_{Z} & = & \left[\bar s_L \gamma^\mu b_L \right] 
%\left[\bar q \gamma_\mu \left ( C_V - C_A \gamma_5\right ) q\right]
%\eqn
%where, $C_V $ and $C_A$ are the vector and axial 
%vector $Z s \bar s $ couplings. Using this operator one can write down
%the effective Hamiltonian for $b \to s s \bar s$ process in this model. 
\beq
{\cal H}^{new}_{Z}&=& -\frac{G_F}{\sqrt{2}} V_{tb}V^*_{ts} 
\left [ \tilde C_3 {\cal O}_3 +  \tilde C_7 {\cal O}_7 +
 \tilde C_9 {\cal O}_9 \right ]
 \label{hamilton_edsq}
\eqn
where, the new Wilson coefficients $\tilde C_3, \tilde C_7$  and $\tilde C_9$ 
at the scale $M_Z$ are given by:
\beq
\tilde C_3(M_Z) &=& \frac{\kappa}{6}  , \\
\tilde C_7(M_Z) &=& \kappa  \frac{2}{3}\sin^2\theta_W, \\
\tilde C_9(M_Z) &=& -\kappa  \frac{2}{3}\left(1- \sin^2\theta_W \right )
\eqn
where, $ \kappa  =  \frac{U_{bs}}{(V_{tb} V^*_{ts})} $, and operators 
${\cal O}_{i}$ in Eq.~(\ref{hamilton_edsq}) are given in Ref.\cite{beneke}. 
We now evolve these new Wilson coefficients from the scale $M_Z$ to the scale 
$\mu \approx {\cal O} (m_b) $ using the renormalization group equation. While doing 
this we have considered NLO QCD correction \cite{buras_rg},
neglecting the order $\alpha $ electroweak contributions 
to the RG evolution equation which are tiny. At the low energy, after the
RG evolution the above three Wilson coefficients 
$(\tilde C_3, \tilde C_7, \tilde C_9 )$  generate new set of Wilson 
coefficients $(\tilde C_i, i = 3-10 )$ in this model. 
The values of Wilson coefficients ( without taking the overall factor 
$\kappa $) at scales $\mu = (m_b/2, m_b) $ are shown in Table 2. 

\begin{table}
\begin{center}
\begin{tabular}{|c|c|c|c|c|c|c|c|c|}
\hline 
scale & $\tilde C_3 $ & $\tilde C_4 $ & $\tilde C_5 $ & $\tilde C_6 $
& $\tilde C_7 $ & $\tilde C_8$  & $ \tilde C_9 $ & $\tilde C_{10}$ \\
\hline
$\mu = m_b/2 $ & 0.195 & -0.088 & 0.0180 & -0.053 & 0.133 & 0.108 & -0.604 
& 0.174  \\
$\mu = mb $ & 0.182 & -0.0629 & 0.0157 & -0.0370 & 0.136 & 0.0732 & -0.574 
& 0.122 \\ 
\hline
\end{tabular}
\caption{Wilson coefficients of EDQS model in NDR scheme, without the overall
multiplicative factor $\kappa $.}
\end{center}
\end{table}
\section{$B$ physics constraints on $U_{bs} $}
In this section, we review the constraints on the flavor violating parameter
$U_{bs}$ from different flavor changing $B$ processes. These processes can
be classified into two classes, CP conserving and CP violating. Among the CP
conserving processes, ${\cal B}(B \to X_s \ell^+\ell^-)$, and $\Delta M_{B_s}$
can put constraints on $U_{bs}$ \cite{ nagashima,botella1,botella2}. 
Using recent Belle \cite{belle_fcnc} 
measurement of ${\cal B} (B \to X_s \ell^+\ell^-) 
= (6.1\pm 1.4^{+1.4}_{-1.1})\times 10^{-6} $ the authors in Ref.\cite{giri}
had shown that $\mid U_{bs}\mid \leq 1\times 10^{-3}$. However, this bound
has recently been updated in Ref.\cite{atwood_hiller} to 
\beq
\mid U_{bs} - 4.0 \times 10^{-4} \mid \leq 8\times 10^{-4}
\label{hiller_bound}
\eqn
which also updates their previous bounds in Ref.\cite{hiller,ali_ball}
The bound in Eq.~(\ref{hiller_bound}) is based on inclusive
$B \to X_s e^+e^-$ decays at NNLO \cite{ali_lunghi}. We shall adopt this 
bound in our analysis.
This bound is valid in both general $n$ extra down-type singlet quark model 
and in the model with a single extra down-type quark singlet~\cite{botella2}. 
Similarly, the $b \to s\gamma $ branching ratio provides comparable 
limits~\cite {chang-chang-keung,botella2}. 

It has been shown in Ref. \cite{jaas,botella3} that in the presence of 
tree-level FCNC coupling $U_{bd} $ and /or $ U_{bs}$, the standard box 
diagram for $B_{d/s} -\bar B_{d/s}$ is not 
gauge invariant by itself, but requires $Z$ exchange penguin diagram as well
as tree level $Z$ FCNC diagram. The additional Feynman diagrams are given in 
Ref. \cite{botella3}. Following the paper \cite{jaas}, with slight change 
in the notations, we write down the expression for the 
$B_q -\bar B_q$ mixing:
\beq
\Delta M_{B_q} = \frac{G^2_F M^2_W f^2_{B_q} {\hat B}_{B_q} m_{B^0_q}}{6 \pi^2}
\left [ \left (\lambda^t_{qb}\right )^2 \eta^{B_s}_{tt} S_0(x_t) + \Delta_{new} \right]
\eqn
where, $(q =d, s )$ and
\beq
\Delta_{new} = - 8 U_{bq} \lambda^t_{qb} \eta^{B_q}_{tt} Y_0(x_t) 
+ \frac{4\pi \sin^2\theta_W}{\alpha} \eta^{B_q}_{Z}U^2_{bq}
\eqn
where, the definitions of different parameters used above can be found in 
Ref.\cite{jaas}. 

We have found that to satisfy the measured $\Delta M_{B_d}$
\footnote{The experimental value for 
$\mid \Delta M_{B_d}\mid = 0.489 \pm 0.008~{\rm ps}^{-1} $ \cite{jaas}.}
within one sigma,
where we consider both the theory error of $20\%$ arising from the value of 
$f^2_{B_q} {\hat B}_{B_q} $ and the experimental error taken in quadrature, 
the FCNC coupling 
$\mid U_{bd}\mid $ should be less than $\sim (2-3) \times 10^{-4}$. This is a 
very stringent limit. 

%Now if there was only a single generation of extra 
%down type quark singlet, then using the Eq.~(\ref{bd_bs_rel}) and the limits
%$\mid U_{sd}\mid \leq 6 \times 10^{-6}$ from $K-\bar K$ mixing and 
%$\mid V_{4b}\mid^2 \leq 0.009$ from the coupling of 
%$Z$ to $b\bar b$ ~[\ref{bbp} - \ref{aguila}, \ref{botella1}] 
%, one can translate the limit on 
%$\mid U_{bd}\mid $ to $\mid U_{bs}\mid \leq (1.8-2.7) \times 10^{-4}$. 
%This bound is well below the value of $U_{bs}$ necessary to explain 
%$ B \to \phi K_S$ data as can be seen later in section 5. Hence, limiting to 
%only one extra down type quark singlet will not explain $ B \to \phi K_S$ data.
%Of course, in the more natural case of arbitrary number of down type 
%quark singlets there is no such constraint on $U_{bs}$ from $U_{bd}$ 
%\cite{botella2}. 

The $\Delta M_{B_s}$ has not been measured yet, and so only
lower limit on the mass difference is available. We have found that the
new contribution to $\Delta M_{B_s}$ from EDQS model is less than $3\%$ 
when compared to the standard model. It can be shown that for similar values of 
$U_{bd}$ and $U_{bs}$, the FCNC effects on  $\Delta M_{B_s}$ will be suppressed by 
a factor $\sim \lambda^2 $ when compared with the effects on 
$\Delta M_{B_d}$. This implies that in EDQS model, FCNC effects are 
hard to detect in $B_s -\bar B_s$ mixing \cite{botella3}.

\section{$B\to \phi K_S$ analysis }
In the last section we have discussed the allowed range of the FCNC parameter
$U_{bs}$ from different $B$ processes. 
In this section we will study the effect of this FCNC parameter $(U_{bs})$ 
in the $B \to \phi K_S$ process. For this 
we express $U_{bs}$ in the following form: $U_{bs} =  \rho e^{i \psi}$.
We will then vary $\rho $ and $\psi$~\footnote{$\psi $ in units of radian}
in range such that 
Eq.~(\ref{hiller_bound}) is satisfied. 
We then study the allowed region of parameters in the $\rho-\psi $ plane 
from the three measured quantities $(a) $ $ {\cal B}(B \to \phi K_S) $, 
$(b)$ $ S_{\phi K_S} $ and $(c)$ $ C_{\phi K_S}$. To get the allowed parameter 
space, from the ${\cal B}(B \to \phi K_S )$ branching ratio, we allow it 
to vary by $2\sigma $ respectively 
from its central value. This $2\sigma$ band contains
both experimental and theoretical errors. The main source of theoretical error 
is the form factor $F^{B \to K}_1$. In our analysis we have considered 
$20 \%$ error on this parameter. Similarly, we vary $C_{\phi K_S}$ and 
$S_{\phi K_S}$ by $1\sigma $ and $2\sigma$ from their central value to get 
the allowed region in the $\rho-\psi $ plane. 

In Figure $2~(a)$ we show such allowed region in $\rho-\psi $ plane for 
the scale $\mu = m_b/2$.
The whole area left of the dotted contour is allowed by saturating 
Eq.~(\ref{hiller_bound}).  
The area outside the thick contour labeled by BR is $2\sigma $ 
allowed region from the branching ratio measurement. 
The parameter space enclosed by the thin contour marked by $S_{\phi K}$ 
is allowed by $2 \sigma $ from data on the $S_{\phi K_S}$. This whole
parameter space is allowed by $1\sigma $ from $C_{\phi K_S}$ measurement, 
The regions (marked by $Z$)  
is the only allowed parameter space in $\rho-\psi$ plane with 
$ 6.5 \times 10^{-4} \leq \rho \leq 10\times 10^{-4}$ and 
$ -1.7 \leq \psi \leq -0.85 $ which satisfy
the experimentally measured $C_{\phi K_S}$, $S_{\phi K_S}$ and 
${\cal B}(B \to \phi K_S) $ within the errors described above. We note that 
only negative values of $\psi $ give acceptable range of $S_{\phi K}$.
The Figure $2~(b)$ correspond to the scale 
$\mu = m_b$. In this case though we have larger allowed area from the 
$S_{\phi K}$ measurement, but the $2\sigma $ branching ratio contour 
pushes the allowed range towards higher values of $\rho $ and somewhat 
lower range of the phase $\psi $. This particular behavior of the branching
ratio contour can be understood from that fact that for $\mu = m_b$, the SM
branching ratio is $ 3.8 \times 10^{-6}$, which is much smaller than the 
lower end of the $2\sigma $ band of the experimental number. Hence, one needs
larger values of $\rho $ to push the total branching ratio within the 
$2\sigma $ limit. For this reason the allowed region shrinks to a point in this
case. 

\section{Conclusions}
In this paper, we have studied the tree-level flavor violating 
$Z$ contribution to $ B\to \phi K_S$ process in models with extra down-type
quark singlets which arise naturally in the context of the Grand Unification 
group $E_6$. In the presence of such flavor violating interactions, 
$B\to \phi K_S $ process receives additional contributions, governed
by a set of new operators which can be expressed in terms of the standard 
operators ${\cal O}_{i}, (i =3-10)$. We then evolved these new Wilson 
coefficients from the scale $M_Z$ to the scale $\mu = {\cal O}(m_b)$ 
relevant for our process using the renormalization group equation. We have 
found that, at the lower scale these Wilson coefficients significantly 
modified from their initial values at the scale $\mu = M_Z$. 
We have found that this new flavor violating interaction can
modify the standard model Wilson coefficients $C_i, (i = 3-10) $
significantly. We have used following experimentally measured quantities: 
$S_{\phi K_S}, C_{\phi K_S}$ and ${\cal B}(B \to \phi K_S)$ to constrain
the flavor violating parameter $U_{bs}$. We have shown that in 
the model with an arbitrary number of down type singlet quarks, the value of
$S_{\phi K_S}$ and $C_{\phi K_S}$ can be well explained by the values of 
$\rho $ and $\psi $ in the region marked by $Z$ in Figure $2~(a)$. Improvements
in measurements of $B \to X_s\ell^+\ell^-$ can tighten the constraints in 
Eq.~(\ref{hiller_bound}) and either rule in or rule out this model. 
\newpage
\begin{flushleft}
\begin{large}
{\bf Acknowledgments}
\end{large}
\end{flushleft}

This work was supported in part by US DOE contract numbers DE-FG03-96ER40969. 
We would like to thank A. K. Giri for helpful correspondence. We would 
particularly like to thank G. Hiller for clarifying the operator structure 
of the $Z$ mediated FCNC interactions.  

\section{APPENDIX : Input parameters and different form factors}

In this Appendix we list all the input parameters, decay constants
and form factors used for the calculation of $B \to \phi K_S $.
\begin{enumerate}
\item {Coupling constants and masses ( in units of GeV ):\\
$ \alpha_{em} = 1/129,~~~~\alpha_{s}(M_Z) = 0.118,~~~~
G_F = 1.16639 \times 10^{-5}~{\rm (GeV)^{-2}},\\
M_Z = 91.19,
~~~~m_b =4.88,~~~~m_B = 5.2787,\\
~~~~m_{\phi} = 1.019,~~~~m_{K} = 0.493 $ }
\item {Wolfenstein parameters :\\
$\lambda = 0.2205,~~~~A = 0.815,~~~~\eta = 0.324,~~~~\rho=0.224,~~~~$
}

\item{Constituent quark masses $m_i (i = u,d,s,c, d )$( in units of GeV):\\
$m_u =0.2,~~~~m_d = 0.2,~~~~m_s = 0.5,~~~~m_c =1.5,~~~~m_b = 4.88. $}
\item{ The decay constants (in units of GeV):\\
$ f_B = 0.19,~~~~f_{\phi} = 0.237,~~~~f_{K} = 0.16 $}
\item The form factors at zero momentum transfer : \\
$F_1^{B\to K} = 0.33. $
\end{enumerate}

%==================================================================%
\def\pr#1,#2 #3 { {Phys.~Rev.}        ~{\bf #1},  #2 (19#3) }
\def\prd#1,#2 #3{ { Phys.~Rev.}       ~{D \bf #1}, #2 (19#3) }
\def\pprd#1,#2 #3{ { Phys.~Rev.}      ~{D \bf #1}, #2 (20#3) }
\def\prl#1,#2 #3{ { Phys.~Rev.~Lett.}  ~{\bf #1},  #2 (19#3) }
\def\pprl#1,#2 #3{ {Phys. Rev. Lett.}   {\bf #1},  #2 (20#3)}
\def\plb#1,#2 #3{ { Phys.~Lett.}       ~{\bf B#1}, #2 (19#3) }
\def\pplb#1,#2 #3{ {Phys. Lett.}        {\bf B#1}, #2 (20#3)}
\def\npb#1,#2 #3{ { Nucl.~Phys.}       ~{\bf B#1}, #2 (19#3) }
\def\pnpb#1,#2 #3{ {Nucl. Phys.}        {\bf B#1}, #2 (20#3)}
\def\prp#1,#2 #3{ { Phys.~Rep.}       ~{\bf #1},  #2 (19#3) }
\def\zpc#1,#2 #3{ { Z.~Phys.}          ~{\bf C#1}, #2 (19#3) }
\def\epj#1,#2 #3{ { Eur.~Phys.~J.}     ~{\bf C#1}, #2 (19#3) }
\def\eepj#1,#2 #3{ { Eur.~Phys.~J.}     ~{\bf C#1},#2 (20#3) }
\def\mpl#1,#2 #3{ { Mod.~Phys.~Lett.}  ~{\bf A#1}, #2 (19#3) }
\def\ijmp#1,#2 #3{{ Int.~J.~Mod.~Phys.}~{\bf A#1}, #2 (19#3) }
\def\ptp#1,#2 #3{ { Prog.~Theor.~Phys.}~{\bf #1},  #2 (19#3) }
\def\jhep#1, #2 #3{ {J. High Energy Phys.} {\bf #1}, #2 (19#3)}
\def\pjhep#1, #2 #3{ {J. High Energy Phys.} {\bf #1}, #2 (20#3)}
%....................................................................%

\newpage
\begin{figure}[hbt]
\centerline{\epsfig{file=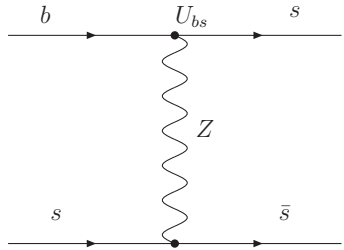,width=\linewidth}}
\vspace*{-15.0cm}
\caption{Feynman diagram for $Z$ exchange tree-level contribution to 
$b \to ss \bar s$ process.}
\label{fig0}
\end{figure}

\begin{figure}[hbt]
\centerline{\epsfig{file=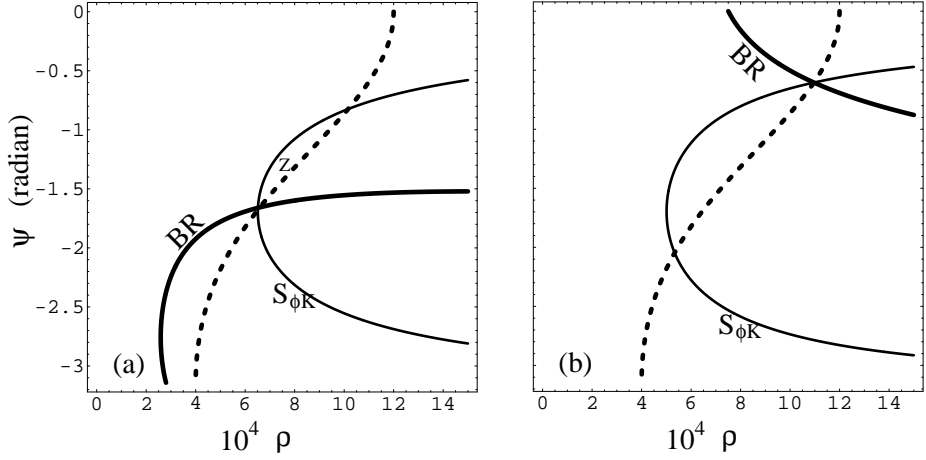,width=\linewidth}}
\vspace*{-8.0cm}
\caption{Contour plots of $S_{\phi K_S} $,  
${\cal B}(B \to \phi K_S)$ and 
$ \mid U_{bs} - 4\times 10^{-4}\mid = 8\times 10^{-4} $ 
in $\rho -\psi$ plane for two 
values of $\mu = \frac{m_b}{2}$ $(a) $ and $m_b$ $(b)$ respectively. 
The area leftside of the dotted contour is the allowed region of
$U_{bs}$ from the inclusive $B \to X_s\ell^+\ell^-$ process. In 
figure $(a)$, the area
marked by $Z$ is the $2\sigma $ allowed regions from the measurement of
$S_{\phi K}$ and ${\cal B}(B \to \phi K_S)$. In figure $(b)$, such 
$2\sigma$ allowed region is a point where the three curves intersect.}
\label{fig1}

\end{figure}

\end{document}